\documentclass[aps,prl,twocolumn,superscriptaddress,amsmath,amssymb]{revtex4-1}
\usepackage{natbib}
\usepackage{graphics}
\usepackage{graphicx}
\usepackage{color}
\usepackage{verbatim}
\usepackage{hyperref}
\usepackage{amsmath}
\usepackage{bm}
\begin{document}

%\preprint{}

\title{Relativistic non-dipole effects in strong-field atomic ionization at moderate intensities}


%\author{}
%\email[]{}
%\thanks{}
\author{Nida Haram}
\email{nida.haram@griffithuni.edu.au}
\affiliation{Centre for Quantum Dynamics, Griffith University, Brisbane, QLD 4111, Australia}

\author{Igor Ivanov}
\email{igorivanov@ibs.re.kr}
\affiliation{Centre for Relativistic Laser Science, Institute for Basic Science, Gwangju-500-712, Republic of Korea}

\author{Han Xu}
\email{h.xu@griffith.edu.au}
\affiliation{Centre for Quantum Dynamics, Griffith University, Brisbane, QLD 4111, Australia}

\author{Kyung Taec Kim}
\affiliation{Centre for Relativistic Laser Science, Institute for Basic Science, Gwangju-500-712, Republic of Korea}
\affiliation{Department of Physics and Photon Science, Gwangju Institute of Science and Technology, Gwangju 61005, Korea}

\author{A. Atia-tul-Noor}
\affiliation{Centre for Quantum Dynamics, Griffith University, Brisbane, QLD 4111, Australia}

\author{U. Satya Sainadh}
\affiliation{Centre for Quantum Dynamics, Griffith University, Brisbane, QLD 4111, Australia}

\author{R. D. Glover}
\affiliation{Centre for Quantum Dynamics, Griffith University, Brisbane, QLD 4111, Australia}

\author{D. Chetty}
\affiliation{Centre for Quantum Dynamics, Griffith University, Brisbane, QLD 4111, Australia}

\author{Igor Litvinyuk}
\email{i.litvinyuk@griffith.edu.au}
\affiliation{Centre for Quantum Dynamics, Griffith University, Brisbane, QLD 4111, Australia}

\author{R.T. Sang}
\email{r.sang@griffith.edu.au}
\affiliation{Centre for Quantum Dynamics, Griffith University, Brisbane, QLD 4111, Australia}


\date{\today}

\begin{abstract}% insert abstract here
We present a detailed experimental and theoretical study on the relativistic non-dipole effects in strong-field atomic ionisation by near-infrared linearly-polarised few-cycle laser pulses in the intensity range $10^{14}$ -$10^{15}$ W/cm$^2$. We record high-resolution photoelectron momentum distributions of argon using a reaction microscope and compare our measurements with a truly ab-initio fully relativistic 3D model based on the time-dependent Dirac equation. We observe counter-intuitive peak shifts of the transverse electron momentum distribution in the direction opposite to that of laser propagation as a function of laser intensity and demonstrate an excellent agreement between experimental results and theoretical predictions.

%{\color{red}whatever}
\end{abstract}

%\pacs{}
%\keywords{}

\maketitle

%\section{Introduction}% Put \label in argument of \section for cross-referencing%\section{\label{}}

In strong-field atomic or molecular ionization with near- and mid-infrared lasers, relativistic effects are known to appear at laser intensities over $10^{18}$ W/cm$^{2}$ \cite{Reiss1998} since, depending on laser wavelength, the photoelectrons can gain sufficiently high ponderomotive energy that their velocity can approach the speed of light in vacuum. However, the onset of relativistic non-dipole effects becomes noticeable even at moderately intense ($10^{13}-10^{14}$ W/cm$^{2}$) low frequency (mid- and near-infrared) laser fields, provided the wavelength is considerably larger than the characteristic atomic size so that the spatial variation of electromagnetic field over the atom cannot be neglected. These non-dipole effects appear as a result of the high energy electrons in the laser field being affected by the magnetic field component of the light pulse, which induces a non-negligible momentum transfer to the photoelectrons \cite{Smeenk2011,Ludwig2014}. To understand and describe these effects, theoretical modelling must extend beyond the dipole approximation, which neglects the small photon momentum by assuming the magnetic field component of the laser field to be zero and resulting in no momentum being transferred to the photoelectrons along the laser propagation direction. In addition, these effects cannot be fully explained without taking into account the Coulomb interaction between the photoelectron and the parent ion, hence going beyond the strong-field approximation (SFA). In case of a linearly polarized laser field, the Coulomb attraction imparts the momentum in the transverse direction and draws the photoelectron towards the parent ion leading to the focusing effect \cite{Rudenko2005,Ivanov2016}. Therefore, the combined effect of Lorentz force and Coulomb attraction from the parent ion contribute to these non-dipole effects. Such non-dipole effects may have a direct implication on many important phenomena relying on re-collision process such as holography with photoelectrons \cite{Huismans2011,Bian2012}, high harmonic generation \cite{Walser2000}, laser induced electron diffraction \cite{Zuo1996,Meckel2008,Blaga2012}, and frustrated tunneling ionization \cite{fti1}.

Relativistic non-dipole effects are revealed in the ionized photoelectron momentum spectra, which are acquired as a result of strong-field atomic ionization. It was reported experimentally that for a circularly polarized laser field (800 nm and 1400 nm, $\sim$ $10^{14}$ W/cm$^{2}$) the ionized photoelectrons gain forward momentum in the laser propagation direction due to radiation pressure effect \cite{Smeenk2011}. The momentum gain is manifested as an asymmtery in the electron momentum distribution (EMD) which can be quantified by extracting a translational shift of the electron momentum spectrum through the measurement of the peak momentum offset. This overall momentum gained from the field is related to the expectation value of the electron momentum in the pulse propagation direction. This value is positive and is well reproduced by the classical \cite{Smeenk2011} and semiclassical \cite{Titi2012} theoretical models. Succesfull attempts were made using theoretical models in the relativistic framework that employs relativistic strong-field approximation (RSFA) \cite{Reiss2013,Yakaboylu2013} and time-dependent Dirac equation \cite{Ivanov2015}. In another experiment; performed with linearly-polarized low-frequency laser field (3400nm, $\sim10^{13}$ W/cm$^{2}$) \cite{Ludwig2014}; the negative peak shifts were reported. A more complex picture emerged when the photoelectron momentum distribution was analyzed in detail. The transverse photoelectron momentum  is shifted forward along the pulse propagation direction only on average. The so-called direct  electrons, which never recollide with the parent ion, are driven in the direction of the laser photon momentum. However, a fraction of slow electrons, which can experience recollision, acquire momentum opposite to the photon momentum. This complex behaviour was shown to be a result of the interplay between the Lorentz force and Coulomb attraction \cite{Chelkowski2015}. Several theoretical models were developed to describe these counter-intuitive offsets, which include semiclassical models \cite{Ludwig2014,Tao2017} and SFA theories such as a non-dipole SFA theory based on exact non-dipole Volkov solutions for TDSE \cite{He2017} and non-dipole quantum trajectory-based Coulomb-corrected SFA theory \cite{Keil2017}. However, proper treatment of such effects clearly necessitates a correct description of both relativistic and Coulomb effects. In \cite{Chelkowski2015} such a description was based on the solution of the time-dependent Schr$\ddot{\text{o}}$dinger equation (TDSE) for a model 2D hydrogen atom described by a soft-core potential. In \cite{Ivanov2016a} these effects were accounted for using the perturbative treatment of the relativistic non-dipole interactions. 

In the present work, we use a theoretical model based on the numerical solution of the 3D time-dependent Dirac equation (3D-TDDE). An approach based on the 3D-TDDE allows a truly ab initio description of the relativistic effects, and has been advocated recently by different groups \cite{Ivanov2015, Kjellsson2017}. Such relativistic development of the theory is indispensable for the description of the ionization phenomena occurring for the currently available field intensities of the order of $10^{18}$ W/cm$^2$ and higher. The approach based on the Dirac equation incorporates relativistic effects, such as the non-dipole effects, effects of the relativistic kinematics and spin-orbit interactions in the most natural way. Comparing to the more traditional approaches, treating the relativistic effects in the framework of the lowest order perturbation theory (LOPT), 
the approach based on the 3D-TDDE offers also an advantage of a technical character which may be quite useful in practical calculations. Indeed, some terms of the Breit-Pauli Hamiltonian \cite{Sobelman72} describing relativistic interactions in the LOPT, are highly singular operators. The spin-orbit interaction operator and the Darwin term, for instance, exhibit at the origin the $1/r^3$ and delta-function singularities, respectively.
Treatment of such singular behavior in a numerical calculation
necessitates the use of various {\it ad hoc} regularization techniques. Solution of the 3D-TDDE, on the other hand, does not pose such technical problems. The price to pay for this advantage is a somewhat higher computational cost which is not, however, an insurmountable obstacle. In the regime of the moderately intense fields considered in this work, we obtain accurate experimental results for the relativistic effects, thereby providing an opportunity for a precise comparison between theory and experiment. 

\begin{figure}[h!]
\includegraphics[scale=0.3]{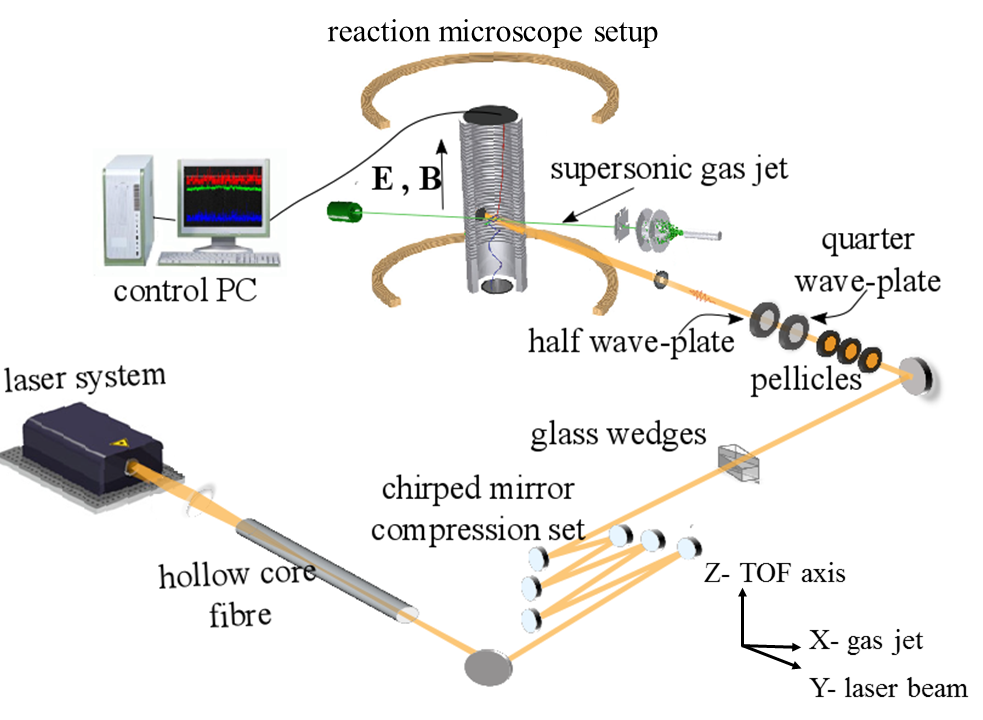}%
\caption{\label{fig:expsetup}}Schematic of the experimental setup
\end{figure}

%\section{Experimental Details}
In our experiment, as shown in Fig. \ref{fig:expsetup}, a linearly polarized few-cycle pulse ($\sim$ 6fs, $\sim$ 740nm, produced by Femtopower Compact Pro CE Phase amplified laser system) propagating along y-axis is focused on a supersonic Ar jet, which is propagating along x-axis. The peak intensity of the few-cycle pulse is varied by a set of pellicle beam-splitters to cover a range of $6\times 10^{14}$-$3\times 10^{15}$ W/cm$^{2}$. A Berek compensator is used to achieve linear polarization of the light pulses, where the inherent ellipticity in the few-cycle pulses is removed by a quarter waveplate. A half wave plate then rotates the polarization axis to z-axis (time-of-flight axis of REMI). A pair of fused silica wedges with adjustable insertion can compensate the chirp of the  few-cycle pulses, minimizing the pulse duration in the interaction region. The in situ laser intensity in the interaction region is precisely calibrated by the recoil-ion momentum imaging method \cite{Smeenk2011a,Alnaser2004} within 10\% confidence. The space charge effect is avoided by carefully controlling the Ar jet density for different laser intensities to achieve a low ionization rate of $<$ 1 ionization event per pulse.

Full 3D photoelectron momentum distributions from the strong-field ionization of Ar atoms is recorded by a reaction microscope (REMI) \cite{ullrich2003}. In order to ensure that the electrons considered are from Ar$^{+}$, both ions and electrons are measured in coincidence with only those electron-ion pairs that mutually conserved momentum.
 
\begin{figure}[!ht]
\includegraphics[scale=0.20]{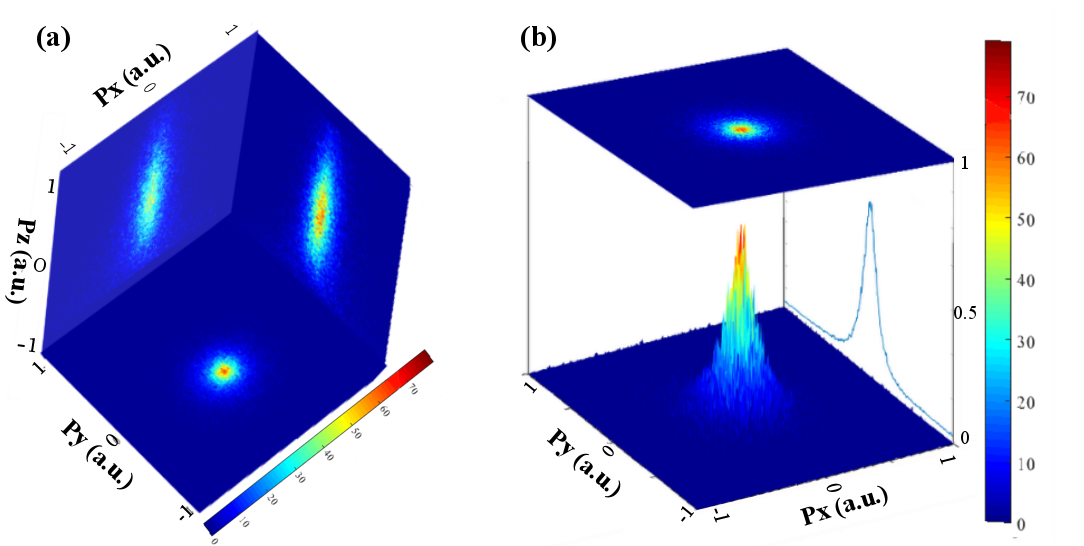}%
\caption{\label{fig:EMD}}(a) Measured electron momentum distributions projected onto three orthogonal planes recorded at 6 x 10$^{14}$ W/cm$^{2}$ (b)Evolution of cusp in the TEMD obtained with an intensity of 6 x 10$^{14}$W/cm$^{2}$
\end{figure}

%\section{Results and discussion}
 Owing to the importance of Coulomb interaction between the parent ion and ejected photoelectron in the linearly polarized laser field \cite{Brabec1996,Corkum1993,Blaga2009,Quan2009,Korneev2012,Liu2012}, the transverse electron momentum distribution (TEMD)--perpendicular to the polarization plane of the laser field--is used to gain insight into the details of recollision event without it being obscured by the large momentum transfer from the laser field. The information about the relativistic non-dipole effects is hidden in the fine details of the TEMD along the laser propagation direction. The TEMD is recorded by a REMI with sufficiently high resolution to resolve the non-dipole effect induced momentum peak shift in the laser propagation direction. The typical 2D photoelectron momentum distributions are presented in Fig. \ref{fig:EMD}(a). The 2D photoelectron momentum spectrum integrated over y-axis evolves into a cusp, as shown in Fig. \ref{fig:EMD}(b). The cusp profile of the TEMD---centred at zero transverse momentum P$_{y}$---is attributed to the Coulomb focusing effect \cite{Brabec1996,Rudenko2005,Comtois2005,Ivanov2016}, which implies that the ionized photoelectron is attracted towards the parent ion due to the momentum transfer in transverse direction as a consequence of Coulomb attraction. 
   
To scale the features of the cusp for a better discernment, the plot of ionization rate $W(P_{y}$) is used. The function $V(P_{y}) = \ln {W(P_{y})}$ is then analyzed in a narrow range of transverse photoelectron momenta $|P_{y}|\le 0.5$ a.u. (a.u. refers to atomic units) \cite{Ivanov2016}. We observed that the cusp profile of the TEMD exhibits slight asymmetry, which becomes more pronounced as the intensity increases. The asymmetry is extracted in the form of peak shift, which is the key observable in this experiment to explore the relativistic non-dipole effects.
For the TEMD to have a cusp profile, $V(P_{y})$ should have a singularity at its maximum near the point$P_{y}=0$, which can be described by representing $V(P_{y})$ in the vicinity of the maximum as  \cite{Ivanov2014,Ivanov2016}:
\begin{equation}
V(P_{y}) = B + A|P_{y} - \beta|^\alpha.
\end{equation}

\begin{figure}[!ht]
\includegraphics[scale=0.28]{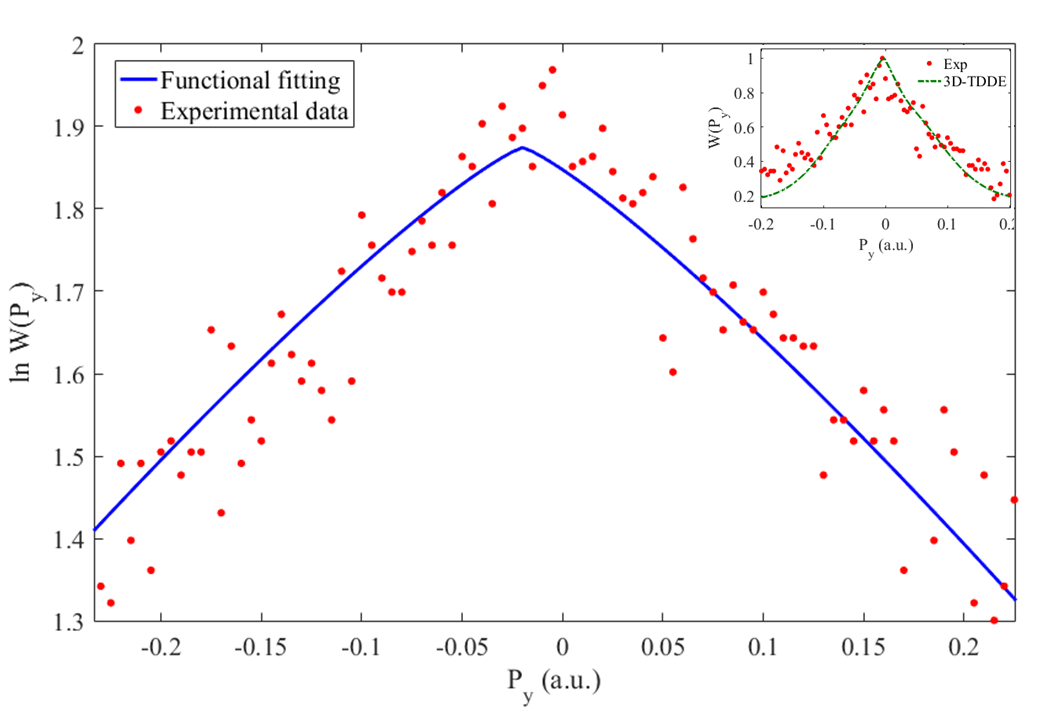}
\caption{\label{fig:fitting}} Projection of TEMD onto the laser beam propagation axis at intensity 6.5 x 10$^{14}$W/cm$^{2}$ fitted with the function $V(P_{y})$ Inset figure shows the normalised ionization rate obtained from theoretical and experimental data
\end{figure}

The experimental data is fitted with the same function $V(P_{y})$ by performing a series of least square fits , as shown in Fig. \ref{fig:fitting}. The coefficients A, B, $\alpha$ and $\beta$ are the fitting parameters. A and B are expansion coefficients, $\alpha$ describes the shape of the TEMD ($\alpha$ $\to$ 1.35 present case) and $\beta$ accounts for the peak shift $\left\langle P_{y}\right\rangle$. The peak shifts $\left\langle P_{y}\right\rangle$ as a function of laser intensity are extracted with reference to the data obtained at the lowest intensity. 

%\section{Theory}
To obtain theoretical predictions, we solve the TDDE:

\begin{equation}
i\frac{\partial \Psi(r,t)}{\partial t}=\hat{H} \Psi(r,t)
\label{d}
\end{equation}

for the bispinor $\Psi(r,t)$ with the Hamiltonian operator: 

\begin{equation}
\hat H = \hat H_{\rm atom}+ \hat H_{\rm int}
\label{h}\ ,
\end{equation}

with:

\begin{equation}
\hat H_{\rm atom}=c{\bm \alpha}\cdot {\hat{\bm p}}+
c^2(\beta-I)+ I\ V(r)\ ,
\label{hatom}
\end{equation}

and

\begin{equation}
\hat H_{\rm int}=c{\bm \alpha}\cdot {\hat {\bm A}} \ ,
\label{hint}
\end{equation}

Here $\bm \alpha$, $\beta$ are Dirac matrices, $c=137.036$ speed of light in atomic units. %${\bm A}$ is the vector potential describing the laser pulse . 

The pulse is defined in terms of vector potential $\bm A (\bm y,t)$ which, when non-dipole effects are included in consideration,  is a function of temporal and spatial variables $A(t-y/c)$ for the pulse propagating in the $y$ direction and is  given as 
\begin{equation}
\bm \hat {A}(\bm y,t)= -\frac{\hat {\bm {E}}}{\omega}\sin^2(\frac{\pi u}{T_1})\cos(\omega u+\phi) \,
\label{hatom1}
\end{equation}
for $0 < u < T_1$, and 0 elsewhere.

Here $u=t-y/c$,  $\bm {E}$ is the field strength, $\omega$ the carrier frequency is $0.059205 a.u.$ corresponding to wavelength 770nm, $T_1=5\times 2\pi/\omega$ corresponds to the pulse duration 6fs  and $\phi$ is the carrier envelope phase (CEP) which is zero in all calculations. 

We use the single-active-electron approximation (SAE) to describe the target atom, employing a model potential $V(r)$ \cite{oep} in Eq. (\ref{hatom1}).

Solution is sought as a series in basis bispinors:

\begin{equation}
\Psi(\mathbf{r},t)=
\sum\limits_{j\atop l=j\pm 1/2} \sum\limits_{M=-j}^{j} 
\Psi_{jlM}({\bm r},t),
\label{basis}
\end{equation}

A detailed description of the procedure used to solve the TDDE numerically can be found in \cite{Ivanov2015}. The photoelectron momentum distribution is obtained by projecting solution of the TDDE after the end of the pulse on the set of the scattering states of the Dirac Hamiltonian (\ref{hatom}) with ingoing boundary conditions.

\begin{figure}[!ht]
\includegraphics[scale=0.36]{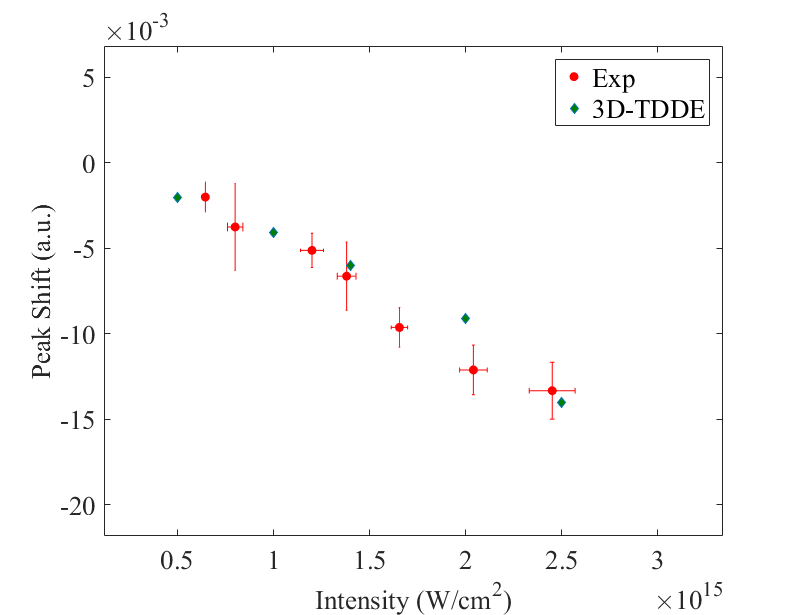}
\caption{\label{fig:exptvstheory}} The relative peak shift $\left\langle P_{y}\right\rangle$ extracted from experiment and theory agree well in the entire intensity range
\end{figure}

In Fig. \ref{fig:exptvstheory}. the experimental results are compared with theoretical predictions with error bars on both intensity and peak shift. The horizontal error bars depict uncertainty in the intensity due to systematic errors, whereas confidence bounds of the fitting parameter $\beta$ serve as error bars on the peak shifts . Clearly, we observe increasing counter-intuitive peak shifts with increasing laser intensities due to the effect of the attractive Coulomb force on the low energy electrons. The experimental and theoretical results are in quite good agreement within the entire intensity range.

In order to unravel the effect of Coulomb potential on the peak shift, we performed simulations for a model Ar atom with Yukawa-type potential:

\begin{equation}
V(r)= -3.3 {e^{-0.3r}\over r} \ .
\label{yukawa}
\end{equation}

\begin{figure}[!ht]
\includegraphics[scale=0.3]{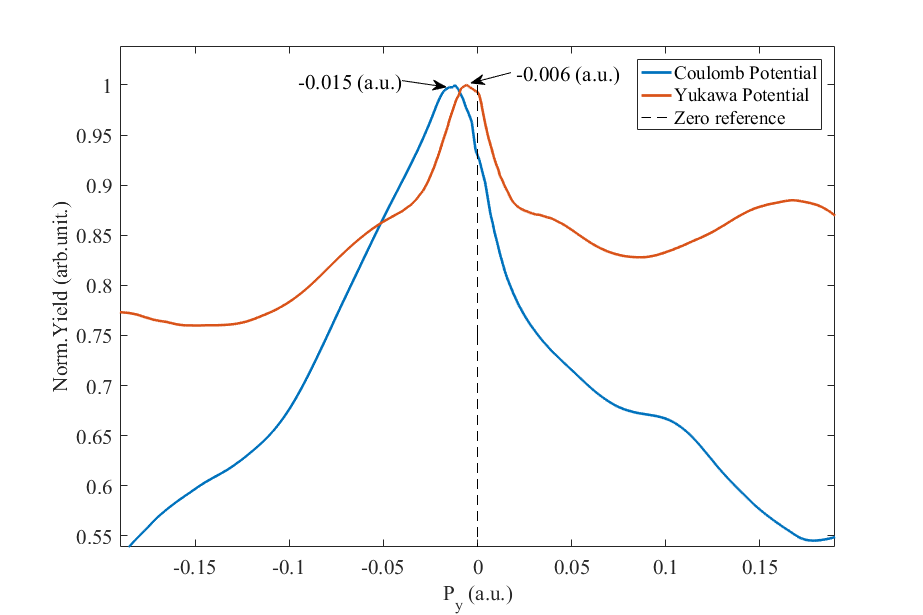}
\caption{\label{fig:dataYukawa}} Simulations with Yukawa potential and Coulomb potential at 2.5x10$^{15}$ W/cm$^{2}$ depicts that Coulomb potential has more pronounced effect on the peak shift compared to Yukawa potential
\end{figure}

The particular values for the parameters of the potential were chosen such that the system had the ionization energy close to that of the Ar atom in the initial $p$-state.  We used initial $p$-state for the ionization in the Yukawa potential, so that all the relevant details of the initial state (energy and angular momentum) would mimic closely those of the Ar atom, to demonstrate the effect of the Coulomb potential unambiguously. The simulations were performed with the same pulses parameters as used in previous simulations at intensity 2.5x10$^{15}$ W/cm$^{2}$, where the effect of Coulomb potential was most pronounced leading to the greatest peak shift. The data with the Yukawa potential (Fig. \ref {fig:dataYukawa}) shows a much less peak shift compared to the data with Coulomb potential, implying that counter-intuitive peak shifts of TEMD  definitely result from the longer range Coulomb attraction of parent ion.

To demonstrate the requirement on the solution of the Dirac equation instead of using the LOPT (for relatively low intensities of the order of $10^{16}$ W/cm$^2$, we compared our theoretical model based on 3D-TDDE with the solution of the time-dependent Schr$\ddot o$dinger equation with leading order relativistic corrections (LOPT-TDSE). The procedure we used for the LOPT TDSE is described in details in \cite{Ivanov2016a}, however the most essential details are detailed as follows:

Our aim is to obtain the leading order relativistic corrections to the non-relativistic TDSE describing evolution of an atomic system. The field-free atomic Hamiltonian including the leading order relativistic corrections, the Breit-Pauli Hamiltonian \cite{Sobelman72}, differs from the non-relativistic Hamiltonian in the terms of the order of $c^{-2}$. If we are interested in the corrections of the order of $c^{-1}$ only, we may, therefore, still use the non-relativistic expression for the field-free atomic Hamiltonian.

For the non-relativistic atom-field interaction Hamiltonian, we use the velocity gauge:
\begin{equation}
\hat H^{\rm nr}_{\rm int}(t) = \hat p\cdot\ {A(t)}+ {\hat A^2(t)\over 2}\,
\label{gauge}   
\end{equation}
where $\bm A (t)$ is the vector potential of the pulse, which in the non-relativistic dipole approximation as a function of the time variable only. Relativistic corrections to this Hamiltonian arise from the fact that vector potential of a traveling wave is a function of time and space variables. In the geometry we employ the laser pulse is polarized along $z$-direction and propagates along the $y$-axis. Vector potential is then a function $\bm A(t-y/c)$ of the combination $t-y/c$. Leading relativistic correction to the operator (\ref{gauge}) can be obtained by substituting this expression for the vector potential instead of $\bm A(t)$ in Eq.(\ref{gauge}), performing expansion in powers of $c^{-1}$ and keeping the terms linear in $c^{-1}$. Following this strategy, we obtain a field-atom interaction operator containing relativistic corrections linear in $c^{-1}$:
\begin{equation}
\hat H^{\rm r}_{\rm int}(\bm r,t) = \hat p_z A(t)+ {\hat p_z y E(t)\over c}+ {A(t)E(t)y\over c} \ ,
\label{gauge1}
\end{equation}
where we introduce electric field of the pulse $\displaystyle E(t)=-{\partial A(t)\over \partial t}$. 

As in the case of the 3D-TDDE we consider Ar atom described by means of a SAE model potential \cite{oep} as a target system. The LOPT-TDSE with the interaction Hamiltonian was solved using the numerical procedure described in \cite{Ivanov2016a}.

\begin{figure}[!ht]
\includegraphics[scale=0.3]{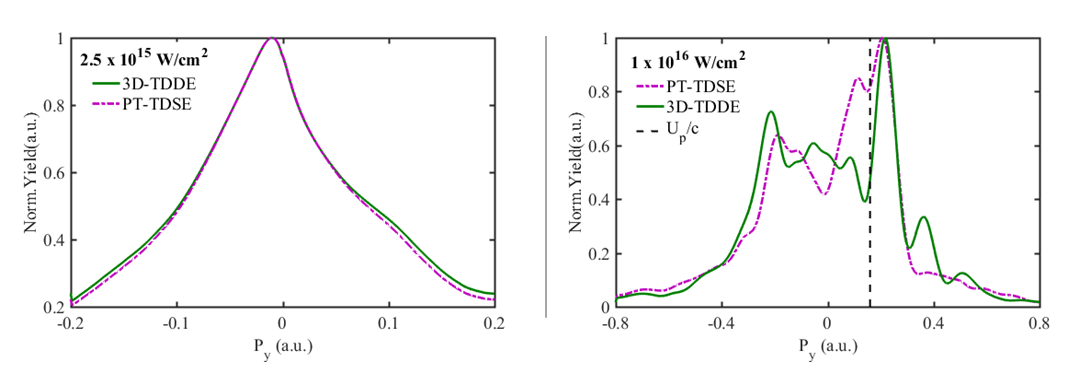}
\caption{\label{fig:PTvsDI}} The TEMDs obtained from the simulations based on 3T-TDDE and LOPT-TDSE at  1 x 10$^{16}$W/cm$^{2}$(right) show quite different features qualitatively compared to the results obtained at lower intensities 2.5 x 10$^{15}$ W/cm$^{2}$ (left)
\end{figure}

The results of the simulations based on 3D-TDDE and LOPT-TDSE show the same peak shift and TEMD features  at 2.5 x 10$^{15}$W/cm$^{2}$, as expected. However, at the higher intensity 1 x 10$^{16}$W/cm$^{2}$, a substantial qualitative difference in the TEMD features was observed even if the peak shift remains the same quantitatively (Fig. \ref {fig:PTvsDI}). It is clear from the right panel of Fig. \ref {fig:PTvsDI}, that at the intensity 
of 10$^{16}$ W/cm$^{2}$
we are entering the regime with the dominating radiation pressure positive peak shift $U_p/c$. The model based on 3D-TDDE offers to probe the fine details in the TEMD and provides information about electron dynamics at high intensities.

Furthermore, the effect of carrier envelope phase (CEP) on the peak shift was investigated experimentally as well as theoretically. It was found that the CEP dependent  peak shift was not resolvable within experimental uncertainty.

%\section{Conclusion}
In summary, we have presented a comparative theoretical and experimental study of the relativistic non-dipole effects emerging at moderately intense laser fields, illustrating the failure of dipole approximation and strong-field approximation even at laser field intensities commonly considered to be non-relativistic. The measured and calculated peak shifts in the direction opposite to the laser beam propagation increase with laser intensity due to the Coulomb attraction from parent ion. Simulations performed for a model Ar atom with Yukawa potential further supports this claim. The truly relativistic ab initio model based on 3D-TDDE agrees quantitatively with the experimental results. The comparison of our model based on 3D-TDDE with LOPT- TDSE indicates that, while the perturbative approach is sufficiently accurate at lower intensities ($<$ 2.5x10$^{15}$ W/cm$^{2}$), there are notable differences in the PEMD simulated using 3D-TDDE and LOPT-TDSE at 1 x 10$^{16}$ W/cm$^{2}$. With 3D-TDDE being the most advanced theory for the description and analysis of relativistic strong-field effects while being only moderately more computationally demanding, its application for a broad range of intensities is fully justified.

%\section*{Acknowledgement}
This research  was supported by the Australian Research Council (ARC)
Discovery Project DP110101894. We also acknowledge support from the Institute for Basic Science, Gwangju, Republic of Korea, under IBS-R012-D1. NH is supported by an Australian Postgraduate Award.
\bibliography{PRL}

\end{document}